\documentclass[conference]{IEEEtran}
\IEEEoverridecommandlockouts
\usepackage{cite}
\usepackage[export]{adjustbox}
\usepackage{graphicx}
\usepackage{multirow}
\usepackage[font={small,it},labelfont=bf]{caption}
\usepackage{xcolor}
\usepackage[utf8]{inputenc}
\usepackage{url}
\usepackage{subcaption}
\usepackage[english]{babel}
\usepackage{booktabs}
\usepackage{blindtext}

\title{Typosquatting 3.0: Characterizing Squatting in Blockchain Naming Systems}
\author{
    \IEEEauthorblockN{Muhammad Muzammil, Zhengyu Wu, Lalith Harisha, Brian Kondracki, Nick Nikiforakis}
    \IEEEauthorblockA{Stony Brook University, New York, USA}
    \IEEEauthorblockA{\{mmuzammil, zhenwu, lharisha, bkondracki, nick\}@cs.stonybrook.edu}
}

\newcommand\bns{BNS}
\newcommand\typo{Typosquatting 3.0}

\newcommand\numTotalENSNames{3,047,188}
\newcommand\numTotalENSNonReadableNames{96,667}
\newcommand\percRecoveryENSNames{97\%}

\newcommand\numTotalUDNames{1,707,017}

\newcommand\numTotalADANames{198,121}
\newcommand\percRecoveryADANames{100\%}

\newcommand\numTotalENSResNames{2,214,012}
\newcommand\numTotalUDResNames{1,156,697}
\newcommand\numTotalADAResNames{198,121}

\newcommand\numTotalENSResolutions{699,548}
\newcommand\numTotalADAResolutions{64,612}
\newcommand\numTotalUDResolutions{393,290}

\newcommand\numTotalENSTxns{140,183,178}
\newcommand\numTotalUDTxns{44,819,348}

\newcommand\numTotalADATxns{15,487,746}

\newcommand\blockXia{13,170,000}

\newcommand\maxTyposENS{102}
\newcommand\maxTyposUD{19}
\newcommand\maxTyposADA{16}

\newcommand\percChangePosAtZero{24\%}

\newcommand\mostTargetsENS{46}
\newcommand\mostTargetsTypos{90}

\newcommand\numComSendersENSCustodial{1,899}
\newcommand\numComSendersPUDCustodial{0}
\newcommand\numComSendersEUDCustodial{38}

\newcommand\numComSendersENS{483}
\newcommand\numComSendersPUD{38}
\newcommand\numComSendersEUD{51}
\newcommand\numComSendersADA{100}

\newcommand\numTwtENSInitList{187}
\newcommand\numTwtAdaInitList{120}

\newcommand\ens{ENS}
\newcommand\ud{UD}

\newcommand\adahandles{ADAH}

\newcommand\escan{Etherscan}
\newcommand\pscan{Polygonscan}

\newcommand\TwtTopENSUD{1,000}

\newcommand\TwtTopADA{100}

\newcommand\TwtTotalENS{14,910}
\newcommand\TwtTotalUD{733}
\newcommand\TwtTotalADA{1,849}

\newcommand\etwtTotalTypos{2,116}
\newcommand\atwtTotalTypos{86}

\newcommand\etwtPercTypos{38\%}
\newcommand\atwtPercTypos{40\%}

\newcommand\etwtDefensiveRegistrations{16}
\newcommand\atwtDefensiveRegistrations{2}

\def\BibTeX{{\rm B\kern-.05em{\sc i\kern-.025em b}\kern-.08em
    T\kern-.1667em\lower.7ex\hbox{E}\kern-.125emX}}
\begin{document}

\maketitle

\begin{abstract}
  A Blockchain Name System (\bns{}) simplifies the process of sending cryptocurrencies by replacing complex cryptographic recipient addresses with human-readable names, making the transactions more convenient. Unfortunately, these names can be susceptible to typosquatting attacks, where attackers can take advantage of user typos by registering typographically similar \bns{} names. Unsuspecting users may accidentally mistype or misinterpret the intended name, resulting in an irreversible transfer of funds to an attacker's address instead of the intended recipient. In this work, we present the first large-scale, intra-\bns{} typosquatting study. To understand the prevalence of typosquatting within \bns{}s, we study three different services (Ethereum Name Service, Unstoppable Domains, and ADAHandles) spanning three blockchains (Ethereum, Polygon, and Cardano), collecting a total of 4.9M BNS names and 200M transactions---the largest dataset for \bns{}s to date. We describe the challenges involved in conducting name-squatting studies on these alternative naming systems, and then perform an in-depth quantitative analysis of our dataset. We find that typosquatters are indeed active on \bns{}s, registering more malicious domains with each passing year. Our analysis reveals that users have sent thousands of transactions to squatters and that squatters target both globally popular BNS domain names as well as the domains owned by popular Twitter/X users. Lastly, we document the complete lack of defenses against typosquatting in custodial and non-custodial wallets and propose straightforward countermeasures that can protect users without relying on third-party services.
\end{abstract}

\date{}

\section{Introduction}
Since the inception of Bitcoin, there has been increased interest in the concept of cryptocurrencies, the blockchains supporting them, and the applications that they enable. Many view the distributed, trustless nature of cryptocurrencies as a welcome alternative to the increased centralization of power and control~\cite{antonopoulos2016internet}. In the context of payments, cryptocurrencies offer willing parties the ability to exchange funds without the need for trusted middlemen that can arbitrarily limit transactions. In the context of the web, the so-called ``second-generation'' blockchains such as Ethereum promise to bring about the next iteration of the web (i.e. Web3 or Web 3.0)~\cite{voshmgir2020token,eth_web3}. This decentralized web allows the deployment of application logic on public blockchains where it can be vetted, ownership of digital assets that are decoupled from any specific third-party service, and the ability of users to manage their own identity by authenticating themselves using their own private keys.

Given that this concept of identity is critical in cryptocurrencies, researchers and developers soon discovered the need to build layers of abstraction on top of the public-key addresses corresponding to each user's wallet. To avoid reintroducing centralization, Blockchain Naming Systems were developed that not only enable the binding of user-friendly strings to wallet addresses (such as \texttt{vitalik.eth} to \texttt{0xd8dA6BF269[...]15D37aA96045}) but store the resolution data on blockchains where only their owner can modify them. Today, modern BNSs not only allow the easier exchange of funds between users but also enable other use cases, such as pointing to web content stored on distributed file-storage networks (e.g. on the InterPlanetary File System~\cite{benet2014ipfs}), resulting in censorship-resistant web applications.

Security researchers have already started studying these new BNS systems, devising threat models and documenting existing types of abuse~\cite{muzammil2024expiredens, xia2022challenges,patsakis2020unravelling,kalodner2015empirical,casino2021unearthing,li2021b,ito2024investigations}. These include hoarding domains for speculation purposes, using takedown-resistant BNS names in the context of malware, domain dropcatching, and squatting trademarks and domains from the traditional web (e.g. attackers owning \texttt{google.eth}).

\begin{figure}[t]
    \begin{center}
        \includegraphics[scale=0.15]{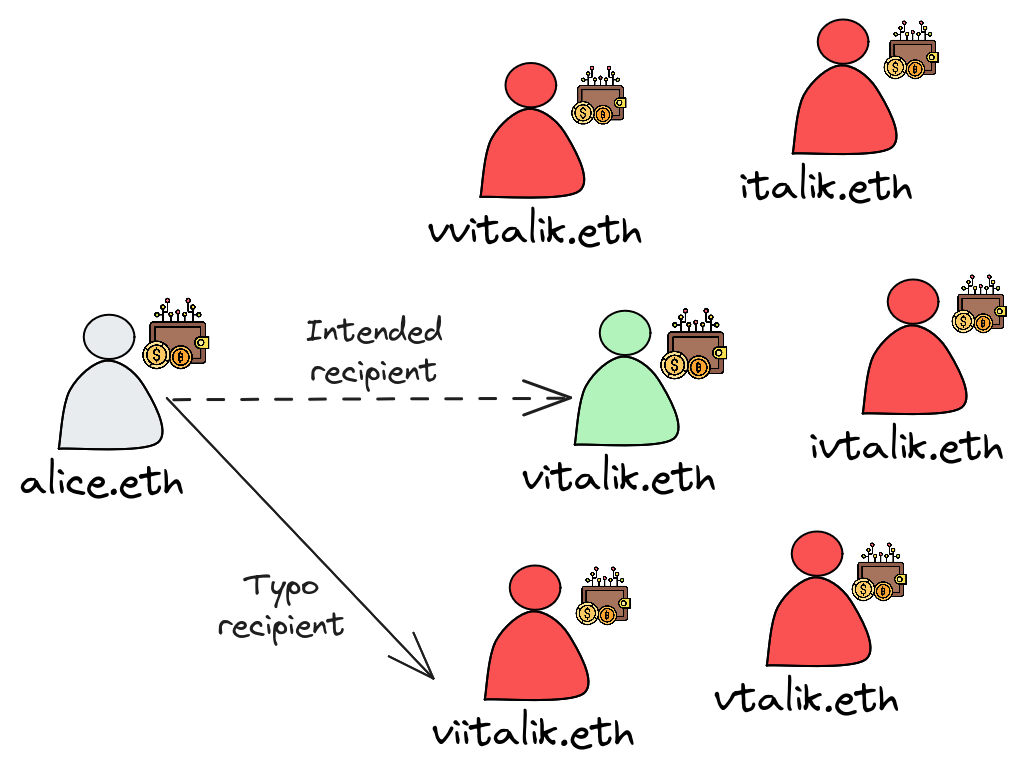} 
    \end{center}    
    \vspace{-3ex}
        \caption{Attackers can ``surround'' benign Web3 domains in order to capitalize from typos. Unlike traditional domain squatting, a single typo can result in the immediate and irrevocable loss of user funds.}\label{fig:typo_example}
            \vspace{-4ex}
\end{figure}

In this paper, we perform the first analysis of \emph{intra-service} typosquatting on popular Blockchain Naming Systems. Instead of looking for which trademarks and domains from the traditional web are being squatted in these BNSs, we focus on attackers registering typo variations of other popular names on the same BNS. This threat model takes into account one of the original uses of BNSs (the exchange of funds between users) and highlights the disproportionate effects of a typo in a BNS, compared to typos in DNS. Whereas a typo in a DNS resolution may require additional social engineering, hosting phishing sites, the downloading of malware, and the exfiltration of sensitive user data, a single BNS typo in the context of a cryptocurrency transaction \emph{guarantees} the loss of user funds. As Figure~\ref{fig:typo_example} shows, all that attackers need to do is register typosquatting variations of popular BNS domains and receive the accidental transactions sent by victim users.

We focus our work on three popular BNSs, namely the Ethereum Name Service (ENS), Unstoppable Domains (UD), and ADA Handles (ADAH). ENS and UD are built on Ethereum (with UD also supporting the minting of domains on Polygon), whereas ADAH is built on top of Cardano. We build a corpus of 4.9 million domain names registered across these BNSs and study the levels of intra-BNS squatting activity, using transaction volume as a proxy for domain popularity. We find tens of thousands of squatting domains across the studied BNSs, targeting as many as 37\% of the most popular legitimate domains. We find that BNS users rarely register typosquatting variations of their own domains, which could route funds mistakenly sent through a typosquatting domain to their own wallet. We report on the makeup of the identified typosquatting domains and show that typosquatting registrations increase year over year, with most squatting domains being registered within 100 days of the legitimate domains they target. 

Next to the domains themselves, we take advantage of the public nature of the three underlying blockchains to understand to what extent attackers have been successful in stealing funds from unsuspecting users. We find thousands of instances where a sender has sent funds to both a legitimate domain \emph{and}  a typo-variation of that same domain, with an average transaction sending \$1,790 to scammers. We confirm the typosquatting phenomenon by focusing on popular cryptocurrency ``influencers'' on Twitter/X and characterize the 74 typosquatting domains targeting the inventor of Ethereum. Lastly, we assess the countermeasures in popular custodial and non-custodial wallets observing a \emph{complete} lack of defenses against typosquatting.

\vspace{0.5ex}
\noindent\textbf{Availability:} To encourage more research in the area of BNS security, our dataset of BNS domains, and scripts to collect blockchain transactions are available here~\cite{artifact_repo}.

\section{Blockchain Name Services (\bns{}s)}~\label{sec:background}
The fundamental concept of \bns{}s (resolving human-readable domains to addresses) is borrowed from the traditional Domain Name System (DNS). However, there are significant differences between the two. Central authorities like ICANN have control over DNS whereas \bns{}s are decentralized naming systems that operate on a blockchain network without a single point of control or failure. Moreover, \bns{} names can be censorship-resistant in that third parties cannot arbitrarily change the resolution address of a BNS domain name.

\noindent\textbf{Ethereum Naming Service (\ens{}).}: \ens{}~\cite{ens} is one of the earliest and the most popular \bns{}, launched in May 2017. It is a decentralized naming system that, at its core, is a collection of smart contracts deployed on the Ethereum blockchain, which uses an account-based transaction model. Smart contracts are self-executing digital contracts with the terms of the agreement directly written into the code. They allow for transactions to occur on the blockchain ensuring anonymity and requiring no form of trust between the sender and the receiver. For example, the Registry contract in \ens{} maintains a list of all domains and subdomains, keeping track of the owners and resolvers for each domain. Resolvers are also smart contracts that respond to queries about a provided domain name, such as returning the wallet address a domain resolves to. Usually, names use the public resolver deployed by \ens{}, but users may develop and use their resolvers tailored to their needs. An \ens{} name is an NFT tied with three main entities: registrant, owner, and controller. The owner of the name is the one who can to change a resolver, expiry date, create or reassign subdomains, or transfer the name to another address, after which the name will change ownership to the new address. The controller may edit the records of a name. The registrant is the owner of a registration~\cite{ensterminology}. \ens{} makes use of a dot-separated hierarchical architecture for its names; names and subdomains registered on it have the extension \texttt{.eth}, as in \texttt{johndoe.eth} and \texttt{funds.johndoe.eth} respectively. A user can register an \ens{} domain for a certain period after which it will expire unless renewed.

\noindent\textbf{Unstoppable Domains (\ud{}).}: \ud{}~\cite{unstoppabledomains} is a \bns{} built on the Ethereum and Polygon blockchains, both following an account-based transaction model. Similar to \ens{} names, \ud{} names are also NFTs controlled by an ``Owner'' and an ``Operator'' (known as ``Controller'' in \ens{}). Both \ens{} and \ud{} provide the functionality of holding a variety of records including cryptocurrency addresses, IPFS hashes, and various text records such as URLs, email, and social-media handles. It is important to note that they can be mapped to addresses for more than one cryptocurrency at a time, allowing users to receive payments in different cryptocurrencies using the same domain name. Some users (either speculators or users/developers who have not yet decided how to use their domain names) may opt not to add any resolution records to their domain names. Unlike ENS, once registered, \ud{} domains do not expire (they can always be sold by their owners to other users). At the time of this writing, \ud{} supports more than ten different TLDs including \texttt{.crypto}, \texttt{.nft}, and \texttt{.blockchain}.

\noindent\textbf{ADA Handles (\adahandles{}).}: \adahandles{}~\cite{adahandle}, based the Cardano blockchain, follows a UTXO-based transaction model. Unlike the previously described \bns{}s, \adahandles{} do not use smart contracts for their operations. Rather, they are stored natively on the Cardano blockchain that supports tokens without the need for smart contracts. This approach circumvents potential issues affecting smart-contract-based BNSs, such as, the need to migrate names to a new smart contract after locating a bug in the current smart contract. ADA Handles are NFTs, and like any other NFT follows a specific minting policy, which is a predefined set of rules governing their creation, distribution, and management. The creators of \adahandles{} publicize the Policy ID they use for this project. An ADA Handle resides inside a wallet address of a user, and payments directed to any handle will be routed to the wallet address in which they reside. Like UD domains, \adahandles{} do not expire and can be owned indefinitely by a user until sold. The use of ADA Handles is growing~\cite{adagrowth} but, being a relatively new naming system on a less popular blockchain, the absolute number of registered domains is smaller than the previous two BNSs. ADA Handles also depart from the traditional hierarchical naming of domain names, instead prepending domains with a dollar sign (e.g. \texttt{\$johndoe}).

\section{Motivation and Challenges}~\label{sec:motivation}
In this section, we introduce the idea and importance of studying typosquatting in BNSs along with similarities and differences with typosquatting in traditional DNS.

\subsubsection*{\typo{}.}
A traditional typosquatting attack in DNS targets users who mistype a domain name in the URL fields of their browsers. In doing so, they give the opportunity to attackers to control the IP address of that resolution and monetize the user's mistake. That monetization typically comes from redirecting users to phishing sites, affiliate-abuse scams, exploit kits, and social-engineering-based attacks. The attacker has to typically convince the user to either provide sensitive information to the landing page (e.g. for phishing sites and survey scams) or willingly accept the download of malware. Even in the worst-case of a resulting malware infection, attackers still need to somehow monetize the infection, either by exfiltrating sensitive data which they can then sell, or relying on ransomware and botnet activities. In short, a typosquatting attack in traditional DNS \emph{may} result in loss of funds, with users having multiple chances to stop the attack before that happens. Even in rarer forms of typosquatting (e.g. users mistyping a recipient's email address) there are still no guarantees that the missent email will contain any sensitive information.

Contrastingly, when a user is trying to send cryptocurrency funds to another user and mistypes the latter's BNS domain name (depicted in Figure~\ref{fig:typo_example}), the loss of funds is direct, immediate, and irrevocable. All that attackers need to do is register typo-variations of popular BNS domains and resolve these domains to their own wallet addresses. There is no need to serve phishing sites, host malware, or in any way try to further engage with victims.

\subsubsection*{Measurement Challenges.} Traditional domain squatting has been extensively studied in past research~\cite{kintis2017hiding, szurdi2014long, agten2015seven, nikiforakis2013bitsquatting, banerjee2008cyber,edelman2003large,moore2010measuring}. It is therefore tempting to assume that all prior methods used by researchers are applicable to studying squatting in BNSs. We argue that effectively studying squatting in BNSs is, in fact, more complicated than studying traditional DNS-based domain squatting for the following reasons:

\begin{figure*}[t]
    \centering
    \includegraphics[scale=0.5]{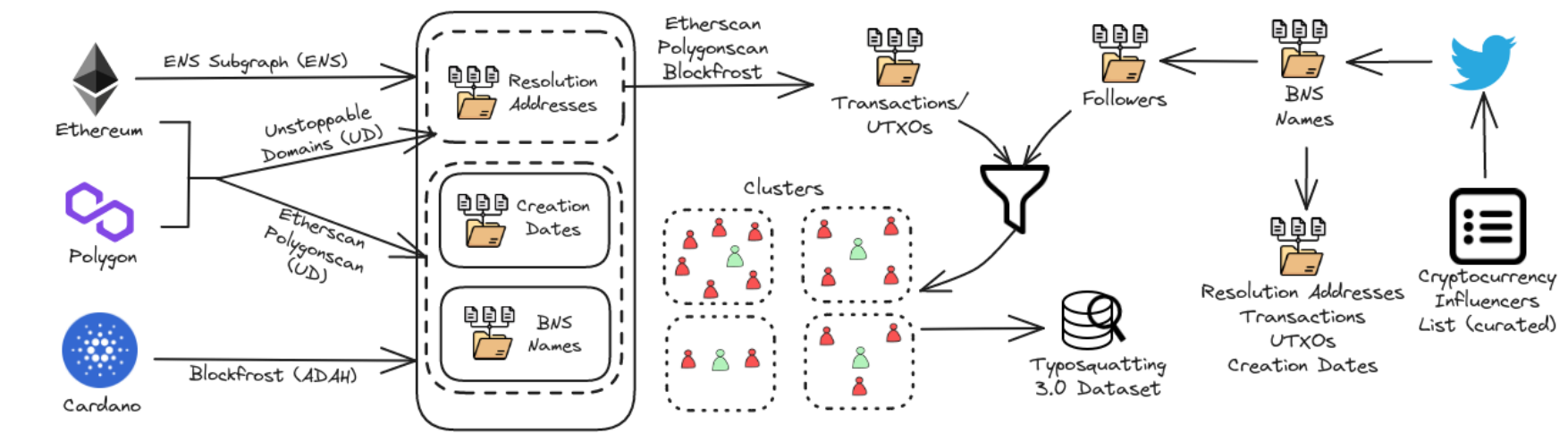}
    \vspace{-1ex}
    \caption{High-level view of our data collection pipeline and how our analysis interfaces with different APIs and third-party services.}~\label{datacollectionpipeline}
\vspace{-3ex}
\end{figure*}

\noindent\textbf{Non-availability of ground truth.} Every domain-squatting study starts with identifying a list of potential targets for attackers. In past studies, given that typosquatting abuse always occurred in the context of navigating the web, popularity of websites was used as a proxy for popularity of domain names. This typically means selecting the domain names of the most popular websites (e.g. top Alexa~\cite{alexa} or top Tranco~\cite{pochat2018tranco}) and mutating them to arrive at possible squatting domains. In BNSs however, domain names are predominantly used in receiving payments with most domain names merely pointing to the wallet address of their owners. As such, looking just at on-chain data related to these domain names, there is no clear way to differentiate popular domains (i.e. domains that squatters are likely to target) from unpopular ones.

  \noindent\textbf{Aliasing of domains to wallets.} Without external sources that can inform the design of a popularity metric for BNS domain names, one reasonable on-chain source of data are transactions. That is, all else being equal, if $addr_1$ has received more incoming transactions than $addr_2$, then it is reasonable to assume that $domain_1$ (i.e. the domain which resolves to that address) is more popular than $domain_2$. The issue with this metric is that it cannot account for multiple domain names resolving to the same wallet address. If $domain_a$ and $domain_b$ both resolve to $addr_1$, there is no on-chain data to conclude which of the two domains is responsible for the most transactions to $addr_1$.

One objective source of off-chain resolution data are software wallets responsible for conducting BNS resolutions. Despite reaching out to multiple wallet vendors with millions of users, those that responded to us informed us that they do not log resolution data and in turn rely on larger platforms for resolutions (e.g. MetaMask relies on Infura for resolutions~\cite{torres2023your}). At the time of this writing, we have not been able to get into contact with these larger platforms and hence must device our own popularity metrics relying on publicly available on-chain data for the majority of our analysis.

\section{Data Collection}~\label{sec:datacollection}
This section describes our data-collection and analysis methodology (shown in Figure~\ref{datacollectionpipeline}) and presents an overview of the datasets used in this work.

\subsection{Collecting Names, Addresses, and Dates}

Given the public nature of the blockchains supporting the evaluated BNSs, all registration/domain-management events are scattered throughout different blockchain blocks. Because of the volume of data available in these chains and the impracticality of linearly searching all these blocks, various third parties (known as ingestion services) mine these blockchains and extract data and insights which they then make available through blockchain explorers. These blockchain explorers can then be used either manually or programatically via traditional web APIs. As much as possible, we rely on these APIs to obtain the BNS domain names, on which the rest of our analysis is built.

\subsubsection*{Ethereum Name Service (\ens{})}
Past work analyzing ENS has shown how difficult it is to extract a complete list of registered ENS domain names directly from the Ethereum blockchain (with or without the use of third-party blockchain explorers)~\cite{xia2022challenges}. This is mostly due to the use of the \emph{namehash} algorithm, allowing ENS to store domains of all lengths as fixed-length (256 bit) cryptographic hashes. Depending on the smart contract mined by researchers, the extraction can be as straightforward as locating the actual domain name in the payload data of a domain-registration event, or as complicated as constructing hashes following the namehash algorithm and then comparing these hashes against the ones controlled by the smart contract.

In 2020, Ethereum developers released ``The Graph'' a decentralized indexing protocol for organizing blockchain data and making that data accessible using GraphQL (a data-query and data-manipulation language used in APIs)~\cite{graphql}. Different organizations can run different ``subgraphs'' for indexing different blockchain data. ENS supports its own subgraph~\cite{subgraph} which sources events from relevant \ens{} contracts. In this paper, we leverage this resource to extract ENS domain names, resulting in a dataset that is \emph{significantly} larger than the one that prior work was able to extract directly from the Ethereum blockchain.

\begin{table*}[]
    \centering
    \scalebox{0.9}{
    \begin{tabular}{@{}|c|c|c|c|c|@{}}
    \hline
    \textbf{BNS}            & \# \textbf{Names Collected} & \# \textbf{Resolvable Names} & \# \textbf{Resolution Addresses} & \# \textbf{Transactions Collected} \\ \hline
    \textbf{\ens{}} & \numTotalENSNames{} & \numTotalENSResNames{}     & \numTotalENSResolutions{}    & \numTotalENSTxns{} \\  \hline
    \textbf{\ud{}}  & \numTotalUDNames{} & \numTotalUDResNames{} & \numTotalUDResolutions{} & \numTotalUDTxns{}  \\  \hline
    \textbf{\adahandles{}} & \numTotalADANames{}   & \numTotalADAResNames{}   & \numTotalADAResolutions  & \numTotalADATxns{}  \\  \hline
    \textit{Total} & \textit{4,962,326} & \textit{3,568,830}                        & \textit{1,157,450}           & \textit{200,490,272}      \\ \hline
    \end{tabular}}
    \caption{Overview of the extracted domains along with the number of resolution addresses and their corresponding transactions.}\label{datasetoverview}
\end{table*}

\subsubsection*{Unstoppable Domains (\ud{})} Due to the lack of a GraphQL endpoint for \ud{}, we adopt an alternative strategy to harvest names. Third parties take advantage of the public nature of popular blockchains to index them, and offer access to their indices via blockchain explorers, such as, Etherscan~\cite{etherscan} and Polygonscan~\cite{polygonscan}. \ud{} names, unlike \ens{} names, are minted on two different blockchains: Ethereum and Polygon. Etherscan and Polygonscan facilitate data retrieval from these respective blockchains, allowing us to query the relevant smart contracts. We focus on querying the smart contracts listed by \ud{} for both Ethereum and Polygon. \ud{} lists separate registry addresses for Ethereum and Polygon minted names, which we utilize to crawl domain names using Etherscan and Polygonscan respectively. Additionally, we leverage another registry smart contract listed on Etherscan that contains a substantial amount of \ud{} names with the \texttt{.crypto} TLD, all minted on the Ethereum blockchain. Using the two block explorers, we query each event log on every block that involved the use of the three mentioned smart contracts. As with any blockchain transaction, the registration of a \ud{} name will also appear as an event log, and will contain the name itself as an argument to the smart contract function. We supplement this data with additional data from the API offered by Unstoppable Domains itself~\cite{udapi} to extract the resolution addresses of each domain name.

\subsubsection*{ADA Handles (\adahandles{})}
To collect \adahandles{} data, we make use of the Blockfrost API~\cite{blockfrost}, an open-source API project that aids in accessing and processing information on the Cardano blockchain, such as addresses, blocks, assets, and transactions. Using the publicly available policy ID for \adahandles{}, we use the ``asset'' endpoint of the Blockfrost API (used to query information regarding NFTs) to collect ADA Handles minted on-chain along with their creation dates and the wallet addresses in which they are located.

\subsection{Obtaining Transactions}

For both ENS and \ud{}, we use \escan{} APIs to collect transactions on the Ethereum blockchain. We provide all the Polygon and Ethereum resolution addresses as inputs to the API calls, upon which they output both incoming and outgoing transactions that each particular resolution address was involved in. Each transaction contains a sender address, receiver address, amount of ETH sent, the transaction hash, and the timestamp. For \ud{} minted on the Polygon blockchain, we use \pscan{} APIs which is queried exactly as the \escan{} APIs.
To collect transactions for \adahandles{}, we use the Blockfrost API to first collect the transaction hashes for each Cardano address in our dataset in which at least one ADA handle resides. For each identified transaction hash, we then collect all associated UTXOs. In the case where we have multiple input addresses in a UTXO for one output address, we count the number of transactions as the number of input addresses and the value as the amount that goes into the output address divided by the number of input addresses.

\subsection{Dataset Overview}

Table~\ref{datasetoverview} provides a summary of the number of domain names that we collected for each \bns{} along with the total number of resolution addresses (i.e. the domain names can be resolved to wallet addresses by the BNS) and transactions. We also report the number of names that have a resolution address. Drawing parallels with traditional DNS, a domain name may be absent from the zone records of an authoritative server for a specific TLD (i.e. it cannot be resolved to an IP address) but it is owned by someone and unavailable for registration.
We note how many \ens{} names have an ``ETH'' resolution address set and how many \ud{} names have an ``ETH'' or a ``MATIC'' address set. Since \adahandles{} resolution addresses are the addresses of the wallets that they reside in, all \adahandles{} had resolution addresses.

We use the described data collection methodology to collect the largest dataset for \bns{}s to date. We obtained \percRecoveryENSNames{} \ens{} names from the ENS Subgraph with only \numTotalENSNonReadableNames{} returning empty API responses. Because of our use of the Subgraph (as opposed to trying to identify ENS domain names through blockchain explorers), we manage to collect 23\% more \texttt{.eth} names than Xia et al.~\cite{xia2022challenges} until block \blockXia{}, which was the cut off for their data collection. Due to API call failures for resolving \ud{}, we were not able to collect resolution address records for approximately 10K \ud{} registered names, which means that we managed to collect over 99\% of \ud{} names that have ever been registered, with \percRecoveryADANames{} recovery rate for \adahandles{}.

Overall, we were able to collect almost 5 million domain names registered across the three evaluated \bns{}s which we analyze in the rest of this paper.

\begin{table}[t]
    \centering
    \scalebox{0.9}{
    \begin{tabular}{|c|c|}
    \hline
    \textbf{Model}         & \textbf{Variation} \\ \hline
    \textit{Duplication}   & jjohndoe.eth       \\ \hline
    \textit{Addition}      & johndoew.eth       \\ \hline
    \textit{Removal}       & johndo.eth         \\ \hline
    \textit{Swapping}      & johnode.eth        \\ \hline
    \textit{Substitution}  & nohndoe.eth        \\ \hline
    \textit{Hyphenation}   & john-doe.eth       \\ \hline
    \textit{Pluralization} & johndoes.eth       \\ \hline
    \end{tabular}}
    \caption{Typosquatting models and variations for \texttt{johndoe.eth}}\label{typosquattingmodels}
    \vspace{-2ex}
\end{table}

\subsection{Establishing Ground Truth}~\label{subsec:groundtruth}
To investigate the prevalence of typosquatting activity using our datasets, it is important to first identify the legitimate names that attackers are likely to be targeting. However, as discussed in Section~\ref{sec:motivation}, establishing this ground truth for \bns{} names is non-trivial because of the unique aspects of BNS domain names, compared to traditional DNS domain names.

We rely on two heuristics to establish ground truth (i.e. a set of legitimate domains that attackers are likely to target via squatting): i) we assume that a resolution address with a significant volume of transactions is owned by a benign user (or at least a user that attackers will target), and ii) a large number of domain names all pointing to the same resolution address constitutes suspicious behavior, suggesting either domain speculators, or squatters who are expecting to capitalize typos when resolving BNS domain names. By combining these two heuristics, we can classify all resolution addresses using the following formula: $w_T / w_D$, where $w_T$ represents the number of transactions linked to each domain's resolved wallet address, and $w_D$ denotes the number of domain names it is linked to. Note that this sorting is based on the transactions of the cryptocurrency resolution address, and not the owner address. Our metric allows us to prioritize the analysis of popular domain names (using the number of transactions as a proxy for their popularity) while penalizing wallets that contain multiple domain names (by uniformly distributing the total number of transactions of that address to all $N$ domains).

\subsection{Cryptocurrency Exchange Addresses and\\Token Contracts}

Through our analysis of the sorted domain names, we discovered a number of BNS names that point to the known wallets of central cryptocurrency exchanges (like Coinbase). Given the popularity of these wallets and their volume of transactions, these domain names ranked near the top of our list. Yet, we can clearly infer that the popularity of these wallet addresses has little to do with the domain names (whether these domain names are managed by the exchanges themselves, or by ENS users who misconfigured their resolution settings is outside the scope of this work). Similarly, we located a number of domain names that resolved to addresses of token contracts as opposed to regular wallet addresses. Given the diversity of these token contracts, we cannot make any definitive general claims about how these domain names are used in the context of transactions.

As a result, we filter any addresses (along with the domain names resolving to these addresses) that belong to known central-exchange wallets and token contracts, as labeled by \escan{} and \pscan{}. Unfortunately, we could not identify a trustworthy source of labeled exchange wallets for Cardano hence we did not filter our \adahandles{} addresses. This lack of filtering is somewhat mitigated by the lower popularity of \adahandles{} which, as shown in Table~\ref{datasetoverview}, has an order of magnitude fewer domain names registered.

\subsection{Identifying Typosquatting Domains}

We define a typosquatting name of a particular \bns{} name as a one belonging to the categories listed in Table~\ref{typosquattingmodels}. These typo models have all been established by prior work in traditional typosquatting~\cite{kintis2017hiding, szurdi2014long, agten2015seven, nikiforakis2013bitsquatting, banerjee2008cyber,edelman2003large,moore2010measuring}.

Given a list of target domain names (which constitute the top $N$ domains after we sort the total list of domain names given our popularity metric), we generate all typosquatting variations using the aforementioned typo models and identify which ones have already been registered. We cluster these typosquatting domains together, according to the domain name that they target. When domain names are short, the likelihood of accidental squatting increases. For example, assuming that \texttt{abc.eth} is a popular ENS domain, is the less popular \texttt{abv.eth} a targeted squatting attack, or a legitimate domain that happens to be syntactically close to the first one? By manually investigating pairs of syntactically-close domain names, we empirically set a lower analysis limit to 5 characters. That is, popular domains that are shorter than that threshold are excluded from the rest of our analysis. Finally, to further reduce false positives, we discard any discovered squatting domains whose registration date predates the corresponding target domain. While these are within a single-character distance of the target domains and may very well end up receiving funds intended for other domains, they do not conform to our threat model of attackers registering domains for the express purpose of squatting existing popular ones.
\section{Analysis}~\label{sec:analysis}
In this section, we discuss our findings we extract through comprehensively analyzing our datasets for each of the three \bns{}s.
\vspace{-2ex}
\subsection{Typosquatting Analysis}~\label{subsec:typoanalysis}

\begin{table}[t]
    \centering
    \scalebox{0.9}{
    \begin{tabular}{@{}|c|c|c|c|@{}}
    \hline
    \textbf{BNS} & \textbf{\begin{tabular}[c]{@{}c@{}}\# Legitimate\\ Names\end{tabular}} & \textbf{\begin{tabular}[c]{@{}c@{}}\# Legitimate\\ Names having\\ atleast one typo\end{tabular}} & \textbf{\begin{tabular}[c]{@{}c@{}}Total\\ Typosquatting\\ Names\end{tabular}} \\ \hline
    \textbf{ENS} & 10,711 & 3,920 (37\%) & 25,396 \\ \hline
    \textbf{UD} & 12,026 & 701 (6\%) & 1,137 \\ \hline
    \textbf{ADAH} & 931 & 191 (21\%) & 396 \\ \hline
    \end{tabular}}
    \caption{Overview for legitimate-to-typosquatting name clusters}
    \vspace{-2ex}
    \label{tab:cluster-overview}
\end{table}

Given that both the Ethereum Name Service (ENS) and Unstoppable Domains (UD) have attracted millions of domain registrations, we choose the top 10K wallets of our ordered list, as the ones that could be targeted by squatters. We refer the domains pointing to these wallets as ``legitimate'' domains, only to differentiate them from typosquatting domains. Given that vast majority of these domains are not attached to known companies and individuals, we cannot be certain that they are associated with legitimate activity. Since ADA Handles (\adahandles{}) have an order of magnitude fewer registrations, we also pick an order of magnitude smaller set of targets (i.e. top 1K domains).

Table~\ref{tab:cluster-overview} provides a high-level overview of the sets of legitimate domains that we selected across all BNSs, along with the number of typosquatting domains targeting them. For both ENS and UD, we observe that there are more legitimate domains than wallet addresses, since a few wallets ``hold'' more than one domains. For \adahandles{} we observe the opposite where some domains were filtered out of our analysis due to their short length. In terms of typosquatting activity, ENS is clearly more targeted than the other two BNSs which is intuitively correct given that it is the oldest and most popular BNS of the three.

\subsubsection*{Number of Typosquatting Registrations.} 
\begin{figure}[t]
    \centering
    \includegraphics[scale=0.5]{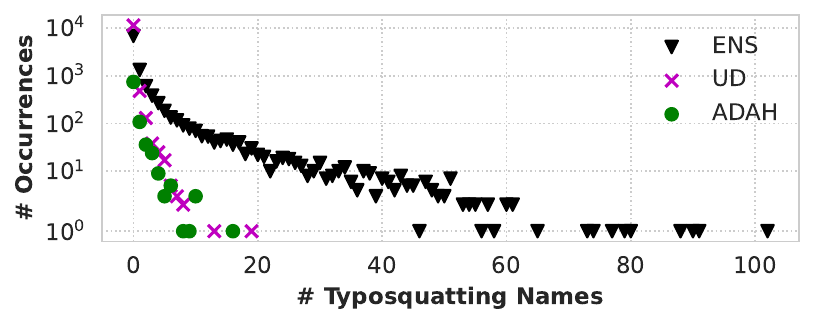} 
    \caption{Frequency of the number of typosquatting registrations against legititmate names}\label{numtypos}
\end{figure}

Figure~\ref{numtypos} shows the number of legitimate names with one or more typosquatting registrations against them. Our analysis reveals that a significant number of names are being targeted more than once. The ENS name targeted the most was \texttt{mickey.eth} with \maxTyposENS{} typosquatting names. Similarly for \ud{}, \texttt{cryptoden.blockchain} attracted \maxTyposUD{} typosquatting names, and \texttt{\$xn\texttt{-{}-}bs8h} for \adahandles{} with \maxTyposADA{} typosquatting names. The ADAH domain is clearly an internationalized domain name (IDN) which corresponds to an emoji when viewed in UTF-8. We assume that not all users can type IDNs in their native format hence we consider their ASCII-based Punycode representations as in scope for squatting.

\begin{figure}[t]
    \centering
    \includegraphics[scale=0.45]{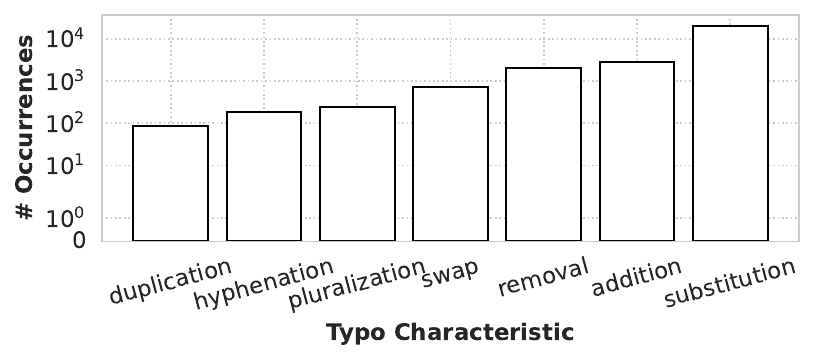} 
    \caption{Frequency of the different misspelling strategies used by typosquatters to form spelling variations of legitimate names}\label{fig3}
    \vspace{-1ex}
\end{figure}

\subsubsection*{Typosquatting Characteristics and Volume of Activity.} Applying different typo models to the same target domain results in tens if not hundreds of possible typosquatting domains. In terms of domains that squatters have chosen to register, Figure~\ref{fig3} presents the models that are most popularly used across all studied BNSs. We observe that the character-substitution model is the one most commonly employed by attackers, followed by character additions, removals, and swaps. In 2015, Agten et al. reported that, in traditional domain squatting, attackers appear to prefer character omission, over introducing additional characters~\cite{agten2015seven}. We also report that \percChangePosAtZero{} of legitimate domains are altered in the first character to produce the typosquatting domain.

\begin{figure}[t]
    \centering
    \includegraphics[scale=0.4]{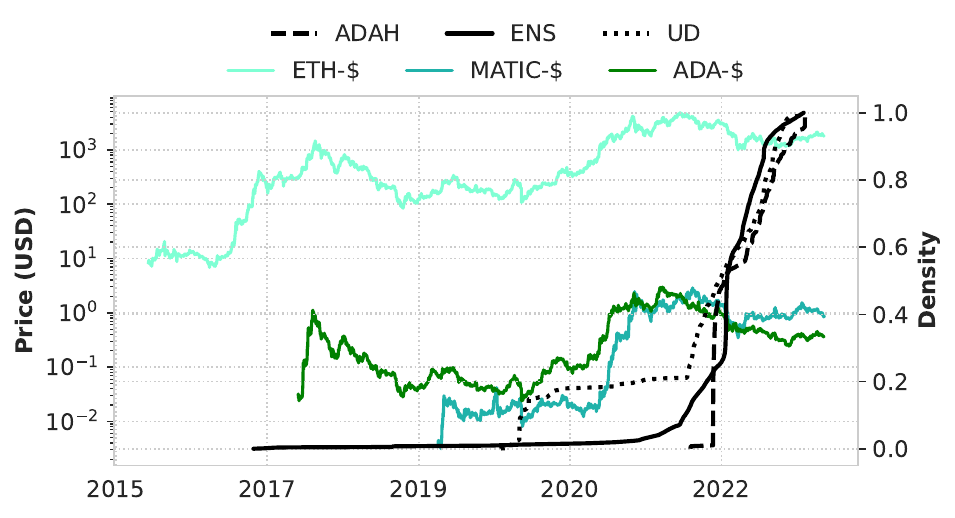} 
    \caption{Correlation between the number of typosquatting names registered each year with the price of each cryptocurrency}\label{prices_registrations}
\end{figure}

Figure~\ref{prices_registrations} shows the fluctuation in prices of Ethereum, Matic, and ADA, along with the number of typosquatting names registered over the years. The trends show that all three \bns{}s have attracted a rapidly increasing number of typosquatting names each year, with a sharp increase in the registration of names after 2021. In 2021, the prices of all three cryptocurrencies reached their maximum exchange rates, which could be a possible explanation for this drastic increase in registrations of typosquatting names. There is no observable slowdown in registrations since 2021 and we expect sustained registrations as cryptocurrencies further gain in popularity.

\subsubsection*{Defensive Registrations.} Like in traditional domain names, users of \bns{}s may proactively register typosquatting variations of their own domain names to protect themselves from the types of attacks discussed in this paper. In the aforementioned 2015 paper, Agten et al. had discovered that 156 of the 500 evaluated top Alexa domains (i.e. 31.2\%)  were engaging in at least one defensive registration of typosquatting domains~\cite{agten2015seven}. In our case, we note a substantially lower rate of defensive registrations, with just 13 ENS legitimate domains (out of the 10,711 selected domains) owned by the same wallet address as at least one typographic variation of itself. We were not able to identify \emph{any} defensive registrations in \ud{} and \adahandles. One possible reason behind this lack of defensive registrations is the current absence of large commercial interest in these BNSs. Companies with billions of dollars in revenue can afford the recurring registrations costs of defensive registrations whereas individuals (even if they are aware of the issue) may not be able to sustain these costs.

\begin{figure}[t]
    \centering
    \includegraphics[scale=0.45]{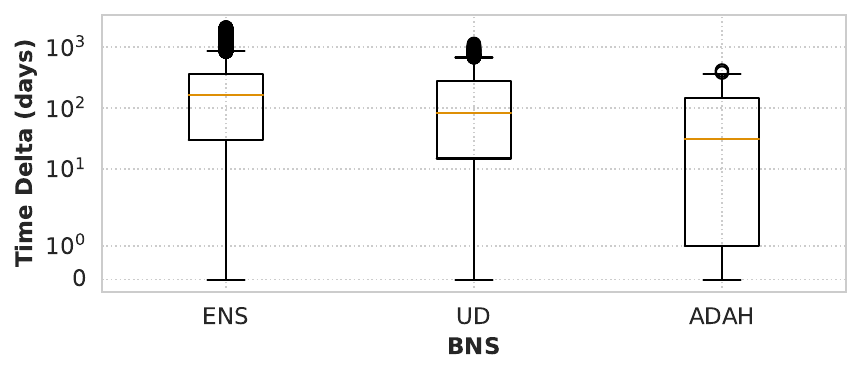} 
    \caption{Time difference (in days) between the registration of the legitimate and typosquatting names}\label{timedelta_regtime}
    \vspace{-1ex}
\end{figure}

\subsubsection*{Trends of Typosquatting Registrations.} Next, we look at the registration-time deltas between the legitimate name and typosquatting names, the results of which are shown in Figure~\ref{timedelta_regtime}. We note that the median time difference is $\approx$100 days, which could be because of the time it takes for a legitimate name to become popular enough to attract squatters. However, it is important to note that the number of typos registered at the very same day of the original name is non zero. One potential explanation for this behavior is that speculators observe the registration of a domain name and predict that it may eventually become popular, thereby immediately registering typo variations of it. One source for these predictions could be social media, where accounts with large followings obtain one or more BNS domain names. Since they are already popular, attackers can immediately register variations of their domain names. We analyze the phenomenon of BNS squatting in relation to social media popularity in Section~\ref{sec:casestudies}.

\begin{figure}[t]
    \centering
    \includegraphics[scale=0.45]{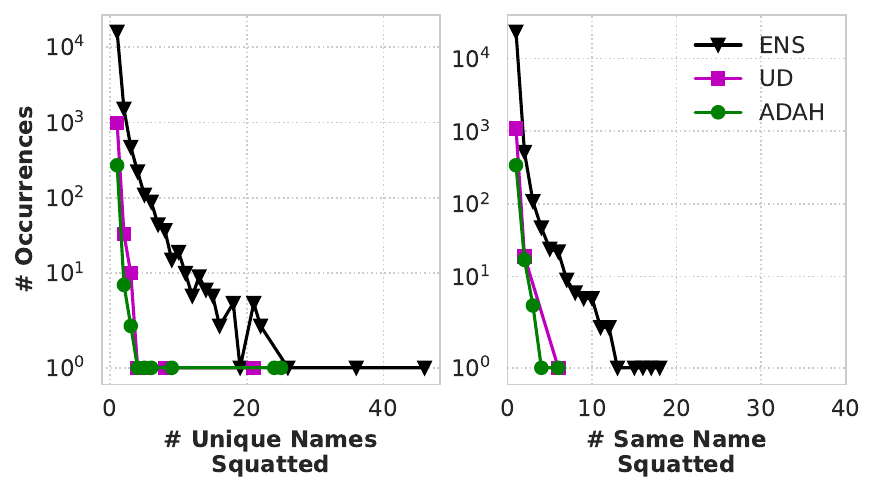} 
    \caption{Number of unique names targeted by the same squatter (Left). Number of times a given name is targeted by a squatter (Right)}\label{squattingvariations}
    \vspace{-3ex}
\end{figure}

\subsubsection*{Behavior of Typosquatters.} 
\begin{figure}[t]
    \centering
    \includegraphics[scale=0.4]{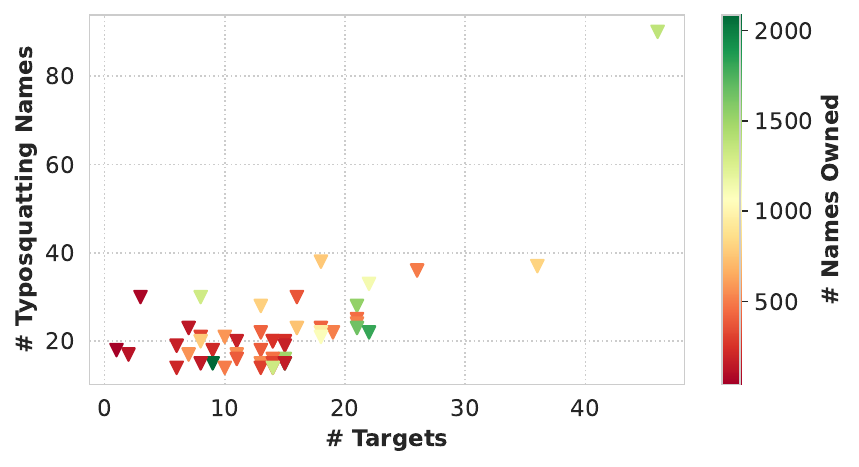} 
    \caption{Number of targets vs. number of typosquatting names for Top 50 \ens{} attackers}\label{ensattackers}
\end{figure}

Furthermore, we explore the behavior of squatters, specifically on how they choose their targets. We analyze the number of times a squatter (as identified by their wallet address) has registered a typosquatting name for the same legitimate name, and the number of unique names a typosquatter has registered a typosquatting name for. Figure~\ref{squattingvariations} shows our results. The attacker that has targeted the most typosquatting names from \ens{} has targeted a total of \mostTargetsENS{} legitimate names, registering a total of \mostTargetsTypos{} typosquatting names, and holds 1,368 domains at the time of our study. Figure~\ref{ensattackers} is focused on the top 50 ENS attackers, in terms of their registration activity. There we see not just that attackers register tens of squatting domains related to the legitimate ones in our study but that they hold hundreds (and occasionally thousands) of other domains in the same wallets. 

\begin{figure}[t]
    \centering
    \includegraphics[scale=0.45]{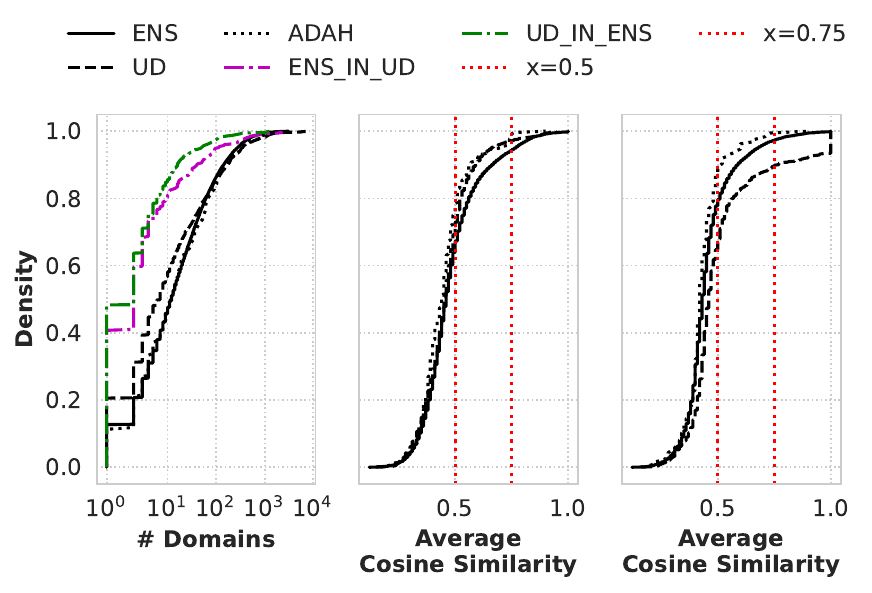} 
    \caption{Distributions for the number of other domains owned by typosquatting wallets (left), the average cosine similarity of typosquatting names and other domains owned by typosquatting wallets (middle), and the average cosine similarity of all domains owned by typosquatting wallets (right)}~\label{numdomains-similarity}
    \vspace{-2ex}
\end{figure}

Of all the typosquatting wallets, 71\% of \ens{}, 87\% of \ud{}, and 66\% of \adahandles{} wallets have registered at least one other domain. Since a wallet address can register both \ud{} and \ens{} names, we also observe that 25\% wallets that had registered \ud{} names, had also registered \ens{} names, and 7\% vice versa, indicating that squatters are registering multiple domains across different \bns{}s as well. The leftmost subplot in Figure~\ref{numdomains-similarity} shows that squatters are indeed registering multiple other domains, indicating suspicious behavior, with the maximum number of domains registered by a single wallet being 3,099 for \ens{}, 6,963 for \ud{}, and 1,477 for \adahandles{}. To further understand the motivation behind this mass registration behavior, we compute the average cosine similarity score of each typosquatting name with all other domains registered by its address (one-to-many), and the average cosine similarity of all names registered by the squatter (many-to-many). Cosine similarity values range from 0 to 1, with a higher score depicting a high semantic similarity. These scores are calculated by using word embeddings obtained using the \texttt{bert-large-uncased} model~\cite{bert}, which is currently state-of-the-art for natural language processing tasks. Using manual inspection, we decide on thresholds to categorize whether two domains are semantically similar in meaning or not, where a score of \(x < 0.5\) indicates that the domains are not similar, \(0.5 \leq x < 0.75\) indicates moderate similarity, and \(x \geq 0.75\) indicates high similarity. Figure~\ref{numdomains-similarity}'s middle and rightmost subplots show our results, where the average cosine similarity trend suggests that names registered by typosquatters are more often than not different in their meaning. From this analysis, we notice that there are three squatting styles among attackers: \textit{i)} registering a single typosquatting domain, \textit{ii)} registering multiple domains for different targets (most popular style of attack in our data), and \textit{iii)} registering multiple domains with similar targets. Examples of each type are shown in Appendix~\ref{appendix}. 

\subsection{Transaction Analysis}~\label{sec:transaction_analysis}
Next, we investigate how successful attackers have been with their typosquatting registrations in \bns{}. Our aim is to identify the number of times users have mistakenly sent funds to an attacker-controlled wallet through a typosquatting \bns{} name. We note that inferring the intent of each transaction is difficult, since on-chain data just reveal the transactions themselves and not which domain the sender used (aliasing issue discussed in Section~\ref{sec:motivation}), let alone which BNS domain a user intended to type. This lack of data can lead to false positives when performing our analysis, as funds going into an attacker-controlled wallet address might not necessarily be through the typosquatting \bns{} name, but rather through another BNS name resolving to the same wallet address. 

To counter these possible false positives, we present two different analyses. First, we report on all transactions that these typosquatting wallets received, with the caveat that some of these transactions may have occurred through domains other than the identified typosquatting ones (i.e. domains that point to the same squatter wallets). Second, we search for senders who have sent funds to both the legitimate \emph{and} the squatting domain, indicating that a mistype was involved in one of their regular transactions. To this end, we extract the transactions of the addresses that resolve from the legitimate name and the corresponding typosquatting names and find the senders that have initiated transactions to both addresses. We report on how many custodial and non-custodial common senders have sent assets to both, the amount of funds that were sent to the squatting address, and the time difference between the transaction to the legitimate address and the squatting ones. We note that the results of this analysis are a lower bound of the transactions that typosquatting domains attracted since we ignore transactions from senders who did not attempt to send any funds to the corresponding legitimate domain. 

On-chain data also lacks relevant insights needed to distinguish whether a transaction was initiated by entering a \bns{} name or the actual wallet address. At the time of our study, only Coinbase, among all custodial wallets, offers a name resolution service, supporting \ens{} and \ud{} domains. This suggests that transactions processed via custodial wallets other than Coinbase most likely occurred through direct wallet address rather than \bns{} name resolution. We exploit this characteristic as a mechanism to filter out such transactions, which accounted to 16\% of all transactions to typosquatting domains.

\begin{figure}[t]
    \centering
    \includegraphics[scale=0.45]{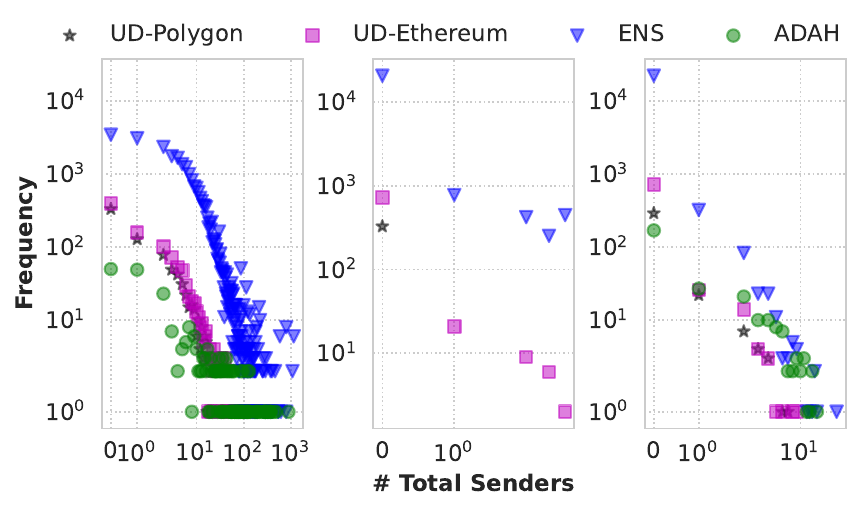} 
    \caption{Frequency of the total number of victims attracted by a unique typosquatting name (left), number of common senders (victims) that have sent assets to a pair of legitimate and typosquatting name through a custodial address (middle)/non-custodial address (right)}\label{num_common_senders}
    \vspace{-3ex}
\end{figure}

\subsubsection*{Victims.} Figure~\ref{num_common_senders} provides three different types of analysis we perform to report the number of transactions towards the typosquatting domains in our dataset. The leftmost figure highlights the frequency of the total number of senders/victims attracted by a typosquatting domain. There are 6 typosquatting \ens{} domains held by a single wallet address which have attracted the highest number of victims, i.e. 1,141. At the time of this study, this wallet holds a total of 112 \ens{} domains that are all either hexadecimal or decimal numbers. Examples that have been flagged as typosquatting domains are \texttt{0x1861.eth}, \texttt{0x1862.eth}, and \texttt{0x1863.eth}, which are targetting the domain \texttt{0x186d.eth} (linked with a wallet that has over 2,000 transactions, and only holds that domain). Similarly, 796 senders have sent assets to 8 different typosquatting names in our dataset that are linked to the same wallet. This wallet owns a total of 99 domains that are all decimal numbers, examples include \texttt{93813.eth} and \texttt{93815.eth} targetting \texttt{95813.eth} (over 5,600 transactions and linked with only one other domain).

Since it is plausible that the transactions were initiated through a domain name not flagged as a typosquatting model but owned by the same wallet, we also report our findings using the common senders approach in the middle and rightmost plots of Figure~\ref{num_common_senders}. In total, we find \numComSendersENS{} pairs of legitimate and typosquatting \ens{} names, \numComSendersADA{} for \adahandles{}, \numComSendersEUD{} for \ud{} on the Ethereum blockchain, and \numComSendersPUD{} for \ud{} on the Polygon blockchain that have received transactions sourced through a non-custodial wallet. In terms of Coinbase (custodial-wallet) sourced transactions, we find \numComSendersENSCustodial{} such pairs for \ens{}, \numComSendersEUDCustodial{} for \ud{} on Ethereum blockchain and \numComSendersPUDCustodial{} for \ud{} on the Polygon blockchain. The top three pairs with most number of transactions from common senders all belonged to \ens{} with the top pair being \texttt{qwerky.eth} and \texttt{qw3rky.eth} (target) receiving transactions from 28 common senders.

\begin{figure}[t]
    \centering
    \includegraphics[scale=0.45]{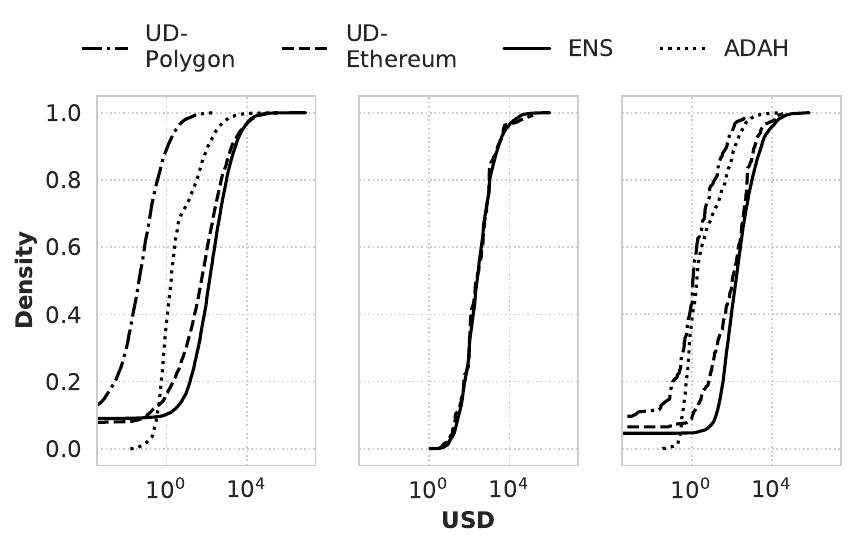}
    \caption{Amount in USD sent to addresses of typosquatting names through the set of all senders (left), common custodial-senders (middle), and common non-custodial senders (right)}\label{funds_sent}
\end{figure}

\subsubsection*{Funds sent to squatters.} Figure~\ref{funds_sent} shows the funds in US dollars that were sent by the set of all senders (left), as well as common senders through custodial and non-custodial wallets to the wallet addresses of the typosquatting names (middle and right). To convert the amount of each cryptocurrency into USD, we utilize the transaction timestamps along with the closing conversion rate for the day of the transaction. Our analysis reveals that the registration of typosquatting names have been successful for attackers in the context of extracting funds from users. From all transactions sent to typosquatting domains, the average amount of funds sent was \$1,790 (\$127 median). The average loss through senders that had sent transactions to both the legitimate and typosquatting names in custodial wallets was \$1,999 (\$241 median) dollars compared to \$1,862 (\$95 median) in non-custodial wallets. This finding supports our distinction between traditional typosquatting and BNS-based typosquatting where a single typo can now cost users thousands of dollars in lost funds.

\begin{figure}[t]
    \centering
    \includegraphics[scale=0.45]{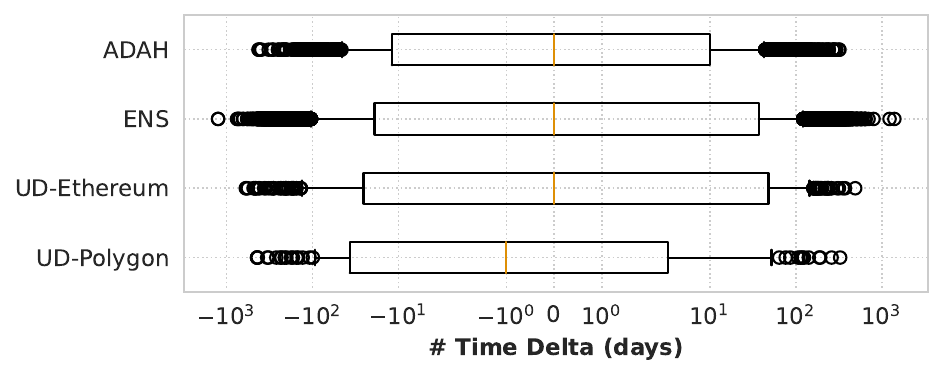}
    \caption{Distribution of time (in days) between a transaction to a legitimate name vs a transaction to a typosquatting name through a common custodial or non-custodial wallet address}\label{time_diff_common_senders}
    \vspace{-2ex}
\end{figure}

\subsubsection*{Differences in transaction times.} Lastly, we calculate the time difference between the transactions of a common sender to a legitimate name vs. its typosquatting equivalent. Figure~\ref{time_diff_common_senders} shows that the median time difference is approximately one day and the interquartile range of all transactions is less than $\pm$100 days around the origin. Both positive and negative differences are intuitive since users may mistype either in their first transaction, or a later one.
\section{Case Studies}~\label{sec:casestudies}
In this section, we complement the high-level results of Section~\ref{sec:analysis} with an analysis of BNS typosquatting activity against popular Twitter/X users, as well as against the inventor of Ethereum.
\vspace{-1ex}

\subsection{Squatting cryptocurrency influencers}
It is common for \bns{} users to publicize their names on their Twitter/X profiles for branding (e.g. associating themselves with a specific technology) or simply to enable their followers to send them cryptocurrency payments. From an attacker's point of view, these Twitter-associated BNS names are ideal typosquatting targets, particularly when their legitimate owners are influencers with millions of followers and thereby a large pool of prospective victims. From our point of view, these influencer-associated BNS names give us an opportunity to verify our earlier observations since Twitter allows us to objectively establish account popularity (and thereby likelihood of targeting specific BNS domains) without restricting ourselves to on-chain data.

\subsubsection*{Data Collection.} We perform our analysis of typosquatting against cryptocurrency influencers as follows: we start by collecting popular influencer accounts in the cryptocurrency space on Twitter. We curated two such lists, one of \numTwtENSInitList{} popular influencers in the cryptocurrency space from the top 3 Google search results in January 2023 for the search query ``Top 100 Cryptocurrency Influencers on Twitter''. Since \adahandles{} were not sufficiently represented in the resulting list, we searched for the keywords ``adahandle'' and ``cardano'' on Twitter and selected \numTwtAdaInitList{} additional legitimate accounts.

Given this initial list of influencer accounts, we use the Twitter API to access the accounts that they follow, and the accounts that their ``followers'' follow in search for other popular accounts that advertise BNS domains in their profiles.

In total, we manage to collect \TwtTotalENS{} \ens{} names, \TwtTotalUD{} \ud{} names, and \TwtTotalADA{} \adahandles{} names from Twitter accounts. In terms of their footprint on social media, \ens{} is significantly more popular than the other two, while \ud{} is the least. We perform another step to narrow down and select the \bns{} names that were found in the profiles of users with large numbers of followers (and thereby large numbers of potential victims for squatters). We create two lists, one comprising of a total of \TwtTopENSUD{} \ens{} and \ud{} domains belonging to popular accounts, and the other consisting of \TwtTopADA{} \adahandles{}s. Our decision to combine the \ens{} and \ud{} domains list was inspired by the fact that very few \ud{} names in our dataset belonged to accounts with significant numbers of followers. Figure~\ref{fig:treysongz} (Appendix~\ref{appendix}) shows the profile of the most popular user on Twitter among those who advertise their ENS domains with 13.1 million followers. Figure~\ref{twtfollowers} (Appendix~\ref{appendix}) shows the number of followers of users collected before and after we filter out unpopular accounts. 
We obtain all relevant addresses and transactions using the same methodology as the one in our Section~\ref{sec:datacollection} analysis.

\subsubsection*{Analysis.} 
\begin{figure}[t]
    \centering
    \includegraphics[scale=0.45]{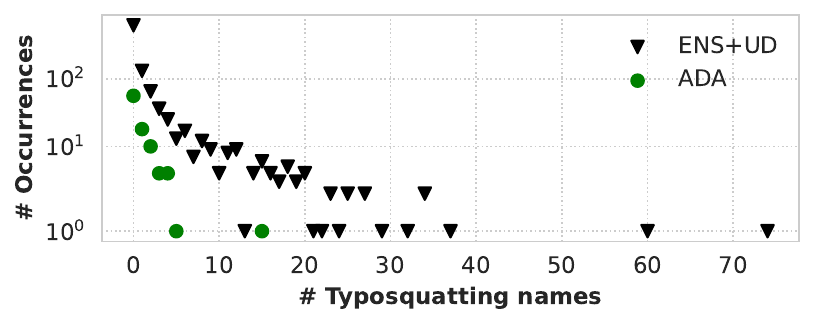}
    \caption{Frequency of the number of typosquatting registrations against legitimate names popularized on Twitter/X}\label{fig:twt_num_typos_each_domain}
\end{figure}

Our findings from analyzing this Twitter dataset suggest that typosquatting activity for such \bns{} names is popular among attackers. Specifically, we indentified \atwtTotalTypos{} typosquatting names related to \adahandles{} and \etwtTotalTypos{} tied to \ens{} and \ud{} names. Of all legitimate \ens{} and \ud{} names, \etwtPercTypos{} were targetted at least once; this percentage was marginally higher (\atwtPercTypos{}) for \adahandles{}. Figure~\ref{fig:twt_num_typos_each_domain} shows the frequency of the number of typosquatting names targeting legitimate names, showing a similar trend as that of Figure~\ref{numtypos}. For example, the \adahandles{}s \texttt{\$bullion} and \texttt{\$villion} are registered against the legitimate name \texttt{\$billion}. A total of 17 names target the \ens{} name \texttt{hayden.eth} (inventor of the Uniswap protocol), and a few examples of typosquatting names include \texttt{hayd3n.eth}, \texttt{haydin.eth}, and \texttt{hyden.eth}. \ud{} domains in the Twitter dataset are overall targeted less than the ones in Section~\ref{sec:analysis}, with some attack instances against popular accounts. For instance, \texttt{metascan.nft} is the \bns{} name owned by the popular Web3 service, MetaScan, and has attracted 3 typosquatting names: \texttt{metascam.nft}, \texttt{metacan.nft}, and \texttt{m3tascan.nft}. 

The three most targeted names in our Twitter dataset from each \bns{} were \texttt{vitalik.eth} (further analysis in Section~\ref{subsection:vitalik}), \texttt{metascan.nft}, and \texttt{\$crypto}. We hypothesise that since such influencers are aware that their \bns{} are well-known, they could be more careful and might have defensive registrations for their domains. However, we find that only \etwtDefensiveRegistrations{} \ens{} users and \atwtDefensiveRegistrations{} \adahandles{} users have one defensive registration each against their \bns{} names. We report that 85 wallet addresses hold more than one typosquatting names that target the legitimate names in our dataset. A single wallet holding 9 domains that are typos of popular domains in our Twitter dataset holds in total over 800 domains.

After performing the same transaction analysis of typosquatting names on this dataset, we find that Twitter-targeting squatters have been just as successful as the ones studied in Section~\ref{sec:analysis}. When all transactions to typosquatting names are considered, an average of \$2,277 per transaction (\$140 median) is transferred to typosquatting names. When focusing just on transactions from senders who have sent funds to both legitimate and typosquatting domains, an average of \$1,192 (\$196 median) per transaction is transferred through Coinbase and \$174 (\$0.5 median) through non-custodial wallets to typosquatting domains. We also find that 297 pairs of legitimate and typosquatting \ens{} and \ud{} names have received at least one transaction through a common custodial sender (Coinbase), and 55 through a common, non-custodial wallet. For \adahandles{}, 8 pairs received funds through the same non-custodial sender.

\subsection{Squatting against Vitalik Buterin}~\label{subsection:vitalik}
In this case study, we report on the typosquatting activity against Vitalik Buterin, the inventor of Ethereum. Vitalik Buterin owns \texttt{vitalik.eth} and advertises it on his Twitter profile. Given the association between Buterin and Ethereum, \texttt{vitalik.eth} is the most squatted domain in our Twitter dataset, with a total of 74 typosquatting names registered against it resolving to 66 unique wallet addresses.

\begin{table}[t]
    \centering
    \scalebox{0.9}{
    \begin{tabular}{@{}|c|c|@{}}
    \hline
    \textbf{\begin{tabular}[c]{@{}c@{}}Typosquatting Name (Number of\\names linked to the same wallet)\end{tabular}} & \textbf{\begin{tabular}[c]{@{}c@{}}Funds (in USD) sent to\\squatter\end{tabular}} \\ \hline
    fitalik.eth (14) & 33,310 \\ \hline
    vitalak.eth (17) & 24,144 \\ \hline
    v1talik.eth (286) & 7,109 \\ \hline
    italik.eth (103) & 3,773 \\ \hline
    vytalik.eth (115) & 1,523 \\ \hline
    \end{tabular}}
    \caption{Typosquatting names that had Coinbase as a common sender from \texttt{vitalik.eth}, and the funds sent to them.}\label{vitalik-short}
    \vspace{-2ex}
\end{table}

\subsubsection*{Typosquatting Name Strategies.} Among the typosquatting names registered, the most popular typo model was that of character substitutions, with 40 typosquatting instances. A few examples of this are \texttt{vitqlik.eth}, \texttt{vlitalik.eth} and, \texttt{vigalik.eth}. Other strategies involved substitutions that convert the name into one that sounds similar (known as soundsquatting~\cite{nikiforakis2014soundsquatting}) including \texttt{witalik.eth}, \texttt{vetalik.eth} and, \texttt{vitalyk.eth}. Less popular strategies included the removal of characters (\texttt{italik.eth}), swapping two characters (\texttt{vitalki.eth}), pluralization (\texttt{vitaliks.eth}), character duplication (\texttt{vitalikk.eth}), and hyphenation (\texttt{v-italik.eth}).

\subsubsection*{Temporal Analysis.} Typos targeting \texttt{vitalik.eth} are constantly being registered starting from just four days after the registration of the original domain (\texttt{vitalikb.eth} was the first typosquatting domain registered in 2017). The name that was registered most recently in our dataset is \texttt{v-italik.eth} in December 2022. The number of typosquatting names registered each year is also increasing, consistent with the overall typosquatting growth observations in Section~\ref{sec:analysis}. Namely, 41 typosquatting domains targeting \texttt{vitalik.eth} were registered in 2022, compared to just 18 in 2021.

\subsubsection*{Transaction Analysis.} Lastly, we perform a transaction analysis, finding that only custodial wallets were used to send funds to \emph{both} \texttt{vitalik.eth} and a typosquatting name. We find 108 transactions where funds have been transferred to typosquatting names using only Coinbase as the source wallet. Table~\ref{vitalik-short} shows the domains these funds were sent to and the amount of funds that has been sent. It highlights not only that the various typosquatting techniques utilized by attackers have been successful, but also that the typosquatters of \texttt{vitalik.eth} have also registered hundreds of other names either hoping to sell them for profit in the future, or capitalizing on typos for multiple accounts. As in all cases involving custodial wallets, we cannot know exactly how many individual users were behind these transactions.

\section{Discussion}~\label{sec:discussion}
In this section we describe the overall limitations of our study and assess the presence of typosquatting countermeasures in wallets and exchanges. Based on our results, we propose some directions for future work in this area.
\subsection{Limitations}

The limitations of this study are grounded on the two challenges listed in Section~\ref{sec:motivation}, i.e., the lack of ground truth regarding domain popularity and the aliasing of multiple domains to the same wallet address. These limitations make it difficult for us to categorically state that each and every transaction sent to a typosquatting variation of a legitimate domain name was the result of a typo. There is insufficient on-chain information to differentiate between a user consciously sending funds to a domain other than a typosquatting domain name, if these two are resolving to the same wallet address. At the same time, we argue that even if some of the identified typosquatting transactions are false positives, this paper sheds light to the overall typosquatting problem in BNSs. Without such a study, centralized exchanges and wallet providers can never try to solve an issue that they do not know they have.

We approached these fundamental limitations through a series of conservative filters, focusing on the senders that have sent funds to pairs of legitimate/typosquatting domain names and thereby excluding one-off transactions that could very well have been the result of typos. Similarly, regarding centralized exchanges, we focused just on Coinbase-originating transactions since that is the only exchange that supports ENS/UD resolutions and thereby the only exchange where transactions could have realistically been the result of a typo. Lastly, our Twitter/X case study confirms the overall problem of typosquatting since we see similar levels of typosquatting activity even when we change our approach for identifying legitimate domains, i.e., domains that attackers target with typosquatting registrations.

\begin{table}[t]
\centering
	\scalebox{0.8}{
    \begin{tabular}{@{}|c|c|ccc|@{}}
    \hline
    \textbf{BNS} & \textbf{Custodial} & \multicolumn{3}{c|}{\textbf{Non-Custodial}} \\ \hline
    \textbf{ENS} &
      Coinbase &
      \multicolumn{1}{c|}{\begin{tabular}[c]{@{}c@{}}Metamask\\ v10.32.0\end{tabular}} &
      \multicolumn{1}{c|}{\begin{tabular}[c]{@{}c@{}}Bitcoin.com\\ v8.5.1\end{tabular}} &
      \begin{tabular}[c]{@{}c@{}}Alpha Wallet\\ v3.65\end{tabular} \\ \hline
    \textbf{UD} &
      Coinbase &
      \multicolumn{1}{c|}{\begin{tabular}[c]{@{}c@{}}Atomic Wallet\\ v1.11.5\end{tabular}} &
      \multicolumn{1}{c|}{\begin{tabular}[c]{@{}c@{}}Bitcoin.com\\ v8.5.1\end{tabular}} &
      \begin{tabular}[c]{@{}c@{}}Alpha Wallet\\ v3.65\end{tabular} \\ \hline
    \textbf{ADAH} &
      N/A &
      \multicolumn{1}{c|}{\begin{tabular}[c]{@{}c@{}}Eternl\\ v1.10.10\end{tabular}} &
      \multicolumn{1}{c|}{\begin{tabular}[c]{@{}c@{}}Nami Wallet\\ v3.5.0\end{tabular}} &
      \begin{tabular}[c]{@{}c@{}}Typhon Wallet\\ v2.5.6\end{tabular} \\ \hline
    \end{tabular}%
    }
    \caption{Digital wallets used in cold/warm typos experiments}\label{tabdefenses}
    \end{table}
\subsection{Defenses and future work}~\label{sec:defenses}
On the traditional web, public DNS servers will resolve all registered domains (both legitimate as well as squatting ones) but some browsing software may warn users about potential typos, requiring confirmation before actually navigating to the website hosted on a suspicious domain~\cite{chrometypos}. One may wonder whether similar countermeasures exist in the systems that support BNS resolutions, namely wallet software for non-custodial wallets and the web applications operated by centralized exchanges.

\subsubsection*{Experiments.} To assess whether these defenses exist, we perform the following two experiments on a centralized exchange that supports BNS resolution as well as on popular wallet software that resolves domains for all three evaluated BNSs:

\noindent\textbf{Cold typos}: Attempt to send funds to a squatting domain without any other related interactions. Do exchanges and wallet software operate global squatting-related blocklists?

\noindent\textbf{Warm typos}: Attempt to send funds to a squatting wallet \emph{after} having sent funds to the legitimate (i.e. squatted) domain. Do exchanges and wallet software operate local (user-specific) squatting-related blocklists?

We use the wallets and custodial exchanges shown in Table~\ref{tabdefenses}. Coinbase is, to our knowledge, the only centralized exchange that supports the resolution of ENS and UD domains at the time of this writing. No exchanges currently support ADA Handles. In terms of non-custodial wallets, we select popular wallets advertised by each respective BNS. For example, the Metamask wallet is installed by more than 10 million users, just on the Google Chrome store. For our cold-typos experiment,  we simply send a minimal amount of cryptocurrency to typosquatting names of popular names from each \bns{} and note the presence of any warnings or errors. For our warm-typos experiment, we first send funds to a legitimate address resolved through a popular BNS domain and then send funds to a typosquatting variant of the same domain.

\subsubsection*{Ethical considerations:} We are familiar with the ethical issues surrounding the sending of funds to attackers, as part of research experiments. For all experiments, we sent the smallest amount of cryptocurrency that the exchange/wallet software would allow us. For example, while Coinbase has a minimum transaction amount of 0.001 ETH, we were able to submit significantly lower transactions through our non-custodial wallets. 

In this, we follow prior security studies where authors make modest payments to attackers and underground economies as a way of shedding light to the studied malicious ecosystems. In past work researchers have, among others, sent small amounts of funds to cyber criminals in order to understand how they construct fake Twitter accounts~\cite{thomas2013trafficking}, operate CAPTCHA farms~\cite{motoyama2010re}, attempt to compromise users~\cite{mirian2019hack}, and create artificial backlinks~\cite{van2019purchased}. As with this prior work, we argue that these small payments are acceptable if the value of the gained insights outweigh the funds sent, and no alternative method was available. 

For this paper, we sent a total of \$12.52 spread over multiple attackers, a vanishingly-small amount considering the actual typo-payments that these attackers are stealing from victims (Section~\ref{sec:transaction_analysis}). These were unavoidable since we lack access to the logic of centralized exchanges and resolution-services of software wallets, meaning that unless we try to send a small payment, we cannot know if there is some resolution-level/transaction-level mechanism that will stop that payment from going through.
\vspace{1ex}
\subsubsection*{Results.} All variations were allowed to go through without any warnings. The only exceptions are Eternl and Bitcoin.com wallets which show a warning after \emph{every} name resolution (for both legitimate and typosquatting domains) reminding users that it is their responsibility to confirm the addresses that they are sending funds to, as all transactions are irreversible. Figure~\ref{croppedssmetamask} (Appendix~\ref{appendix}) illustrates the screen views from Metamask when sending funds to a typosquatting domain. These results highlight that there is ample room for deploying global and local defenses to protect cryptocurrency users. Even without any trusted third parties (i.e. along the overall ethos of cryptocurrencies and trustless P2P payments), software wallets can keep local lists of the domains that their users have resolved and sent funds to, calculate typosquatting variations using well-known typo models, and warn users if they ever try to send funds to these variations in the future.

\section{Related Work}
~\label{sec:relatedwork}
To the best of our knowledge, this is the first study that investigates the issue of intra-squatting in modern BNSs, i.e., users of a BNS targeting other users of the same BNS. In this section, we briefly describe the limited related work in the BNS space, as well as how our work compares to squatting research in traditional DNS domain names.

The work that is most closely related to ours was the \ens{} measurement study by Xia et al.~\cite{xia2022challenges} in which the authors present a systematic analysis of the Ethereum Name Service, its growth, and the different types of attackers that it has attracted. In terms of squatting, the authors focus on \emph{inter-system} squatting, i.e., evaluating to what extent attackers are squatting on popular domains and trademarks that do not belong to them (e.g. \texttt{google.eth} and \texttt{facebook.eth}). The threat models behind these types of squatting domains are distinctly different from the ones we explored in this paper with attackers hosting malware or intending to sell the squatting domains back to their original trademark holders, as opposed to monetizing typos occurring during the exchange of funds. Similar to that work, Patsakis et al.~\cite{patsakis2020unravelling} describe possible squatting attacks on decentralized blockchain-based naming services focusing on NameCoin and EmerCoin, two blockchain-based name services that behave like traditional DNS, resolving domain names to IP addresses. The authors evaluate the level of trademark squatting in these two services, but do not explore intra-chain squatting, which is the focus of this paper. Kalodner et al.~\cite{kalodner2015empirical} explore the prevelance of name-hoarding in NameCoin's \texttt{.bit} TLD domains, where users purchase domains with the intent of speculative resale at higher prices. They develop techniques to analyze the transfers of these domains from one user to another and find that at the time of their study, transfers of these domains was rare. Muzammil et. al~\cite{muzammil2024expiredens} investigated dropcatching attacks on ENS domains, where attackers can re-register expired ENS domains to attract transactions that were meant for their previous owners. Apart from BNS attacks, various types of scam activities in the Web3 space has caused significant amounts of financial losses~\cite{li2023double, li2024like, liu2024give, li2023towards, na2023evolving, boshmaf2020investigating,kim2023drainclog,das2022understanding, xu2019anatomy, phillips2020tracing, abramova2021bits, huang2018tracking}.
\vspace{1ex}

In traditional DNS, cybersquatting (individuals registering trademarks not belonging to them) and typosquatting (registering typo-variations of existing popular domains) can be traced back to the 1990s~\cite{acpa, kesmodel2008domain}. Due to their popularity and level of abuse, these phenomena attracted a large body of research attempting to understand how typos are constructed, how fast users get compromised, whether keyboard layouts affect typosquatting, and how typosquatting domains are abused by their owners~\cite{le2019smorgaasbord, khan2015every, szurdi2014long,agten2015seven,edelman2003large,banerjee2008cyber,moore2010measuring, tahir2018s, koide2023phishreplicant, blefari2024typoalert, valentim2024x, majumdar2024beyond, lepipas2024username}. Researchers also identified and studied other types of squatting including \emph{homograph domains} (malicious domains that abuse visually-similar characters with their victim domains~\cite{holgers2006cutting}), \emph{soundsquatting} (malicious domains that sound like popular domains~\cite{nikiforakis2014soundsquatting, valentim2022ai}), \emph{bitsquatting} (malicious domains that are one-bit-flip variations of popular domains~\cite{nikiforakis2013bitsquatting, dinaburg2011bitsquatting}), and \emph{combosquatting} (malicious domains that include popular trademarks~\cite{kintis2017hiding, vu2020typosquatting}).
\vspace{1ex}

The main difference between the BNS squatting we studied in this paper and all types of traditional domain squatting is the complexity and likelihood of successfully exploiting users through a squatting attack. As we argued in Section~\ref{sec:motivation}, BNS squatting is a \emph{significantly} stronger attack vector where a single typo in a wallet software or online exchange translates to the immediate and irrevocable loss of funds, without attackers needing to further social-engineer users, infect them with malware, or exfiltrate sensitive information from their machines.
\section{Conclusion}~\label{sec:conclusion}
In this paper, we drew attention to the issue of typosquatting in Blockchain-based Naming Systems (BNSs) and performed the first study of intra-chain squatting across three \bns{}s. We were able to build a corpus of 4.9 million domains hosted across these services which we used to define legitimate domains and identify squatting domains targeting them. Through a set of conservative filters aimed at  overcoming the inherent limitations of the available on-chain data on these domains, we discovered tens of thousands of squatting domains, targeting as many as 37\% of the benign domains in their corresponding BNS. Among others, we observed an increasing number of typosquatting registrations in BNSs per year, with defensive domain registrations being almost entirely absent. In terms of transactions, we focused on senders who have sent funds to pairs of legitimate/typosquattting domains, observing thousands of such pairs with transactions involving substantial amounts of cryptocurrencies, often the equivalent of hundreds to thousands of dollars. Lastly, we confirmed the typosquatting phenomenon on Twitter data and observed that modern software wallets and online exchanges do not currently support any mechanisms to protect their users from typosquatting.

\vspace{1ex}
\noindent\textbf{Acknowledgements} We thank the anonymous reviewers for their helpful feedback. This work was supported by the National Science Foundation (NSF) under grants CNS-2211575, CNS-2126654, and CNS-1941617.
\bibliographystyle{unsrt}
\bibliography{paper.bib}
\begin{appendices}
\section{}~\label{appendix}
\begin{table*}[h]
  \centering
  \scalebox{1}{
  \begin{tabular}{|c|c|c|}
  \hline
  \multirow{2}{*}{\textbf{Target}} & \textbf{Typosquatting} & \textbf{Typosquatting}  \\
& \textbf{Name} & \textbf{Address} \\ \hline
  vitalik.eth      & vitqlik.eth                 & 0x6c30...83a90 \\ \hline
  vitalik.eth      & vita1ik.eth                 & 0x4bcb..d920 \\ \hline
  vitalik.eth      & vitlaik.eth                 & 0x90e9...84d6 \\ \hline
  vitalik.eth      & vitalii.eth                 & 0x4c81...5492 \\ \hline
  vitalik.eth      & vitaiik.eth                 & 0x0325...e09ec \\ \hline
  vitalik.eth      & vitalki.eth                 & 0x8c26...fe68 \\ \hline
  vitalik.eth      & vitalij.eth                 & 0x5201...80a3 \\ \hline
  metascan.nft     & m3tascan.nft                & 0xcd9f...aab9 \\ \hline
  play2earn.crypto & play2ern.crypto             & 0xaf81...b4bb \\ \hline
  jjlin.eth        & jhlin.eth                   & 0xc3e0...67e7 \\ \hline
  jjlin.eth        & jnlin.eth                   & 0x3682...333a \\ \hline
  jjlin.eth        & jlin.eth                    & 0x793b...0805 \\ \hline
  blackcoin.crypto & balckcoin.crypto            & 0x8e77...0fa4 \\ \hline
  \$ada\_astronaut & \$adaastronaut              & addr1q82...40t9                \\ \hline
  \$cnftjunky      & \$cnftjunk                  & addr1q9t...uq6c               \\ \hline
  \end{tabular}}
  \caption{Examples of targets, typosquatting names, and typosquatting addresses of \textbf{Type (i)} typosquatting attacks (i.e. where attackers register only one typosquatting domain, and no other domains)}~\label{tab:camp1}
\end{table*}

\begin{table*}[h]
  \centering
  \scalebox{1}{
  \begin{tabular}{|c|c|c|c|c|c|}
  \hline
  \textbf{Address} &
    \textbf{0x4ea8...567e} &
    \textbf{0x7afd...e4a2} &
    \textbf{0xba44...ae33} &
    \textbf{addr1q87...py2f} &
    \textbf{addr1qyd...hk6z} \\ \hline
  \multirow{18}{*}{\textbf{\shortstack{Sample\\Domains}}} &
    googleeth.eth &
    vitalik3.eth &
    0ldtrafford.crypto &
    \$-handles- &
    \$olidity \\ \cline{2-6} 
   & treysong.eth       & masterkard.eth    & 0ldtrafford.nft       & \$-handle-         & \$occer           \\ \cline{2-6} 
   & ahopkin.et         & maricle.eth       & vitalybuterin.nft     & \$-handles         & \$tewart          \\ \cline{2-6} 
   & vitulik.eth        & mckillip.eth      & vitalik0.x            & \$-handle          & \$anchez          \\ \cline{2-6} 
   & vitalikbuteri.eth  & bl0ckhead.eth     & j0ebiden.nft          & \$dana.swap        & \$hrek            \\ \cline{2-6} 
   & votalik.eth        & btcafe.eth        & b0bmarley.nft,        & \$dana\_           & \$stevenshandle   \\ \cline{2-6} 
   & paradigmi.eth      & akward.eth        & 0x00001.x             & \$bitfinex         & \$steveshandle    \\ \cline{2-6} 
   & paradigmu.eth      & beelive.eth       & 0x000001.x            & \$binance          & \$paulshandle     \\ \cline{2-6} 
   & pradigm.eth        & greatfull.eth     & k0bebryant.blockchain & \$usbank           & \$tephanieshandle \\ \cline{2-6} 
   & jimmyfalloneth.eth & dallasfans.eth    & krypt0.blockchain     & \$-eternlwallet-   & \$stephanies      \\ \cline{2-6} 
   & starbuk.eth        & chesleafans.et    & 0ptimusprime.nft      & \$-eternl.wallet   & \$stephans        \\ \cline{2-6} 
   & mariogotz.eth      & heatfan.eth       & an0nymous.blockchain  & \$-eternl\_wallet- & \$tephanshandle   \\ \cline{2-6} 
   & pornhb.eth         & wilsonfamily.eth  & 0xpunk.crypto         & \$ada\_handles-    & \$ex-appeal       \\ \cline{2-6} 
   & pornhbu.eth        & taylorfamily.eth  & nakamoto1.crypto      & \$ada\_handles.    & \$ex\_appeal      \\ \cline{2-6} 
   & simsung.eth        & figurati.eth      & stevej0bs.crypto      & \$adahandles       & \$sex.appeal      \\ \cline{2-6} 
   & amsang.eth         & johnwaynegacy.eth & w0nderw0man.nft       & \$ada-handle-      & \$haman           \\ \cline{2-6} 
   & davidsiwonchi.eth  & p0rntv.eth        & daviddunn.nft         & \$walgreens        & \$motor-yacht     \\ \cline{2-6} 
   & davidsiwonchio.eth & nfasset.eth       & w0lfman.nft           & \$rockwell         & \$showbiz         \\ \hline
  \textbf{\begin{tabular}[c]{@{}c@{}}Total\\ Domains\\ Owned\end{tabular}} &
    116 &
    773 &
    5862 &
    1475 &
    438 \\ \hline
  \end{tabular}}
  \caption{Examples of targets, typosquatting names, and typosquatting addresses of \textbf{Type (ii)} typosquatting attacks (i.e. where attackers register multiple domains targetting dissimilar legitimate domains)}~\label{tab:camp2}
  \end{table*}

\begin{table*}[h]
  \centering
  \scalebox{1}{
  \begin{tabular}{|c|c|c|c|c|c|}
  \hline
  \textbf{Address} &
    \textbf{0xc900...e268} &
    \textbf{0x250a...cc2f} &
    \textbf{0xc7d9...2434} &
    \textbf{0x143a...a1d1} &
    \textbf{0x92c3...bb57} \\ \hline
  \multirow{6}{*}{\textbf{\shortstack{Sample\\Domains}}} &
    w3bank.eth &
    98223.eth &
    derick2.eth &
    0878888888.blockchain &
    1818181818.wallet \\ \cline{2-6} 
    & web3nk.eth & 98454.eth & erick2.eth & 0908888888.888     & 181818181818.wallet     \\ \cline{2-6} 
    & web4nk.eth & 98534.eth &            & 0998888888.888     & 1818181818.blockchain   \\ \cline{2-6} 
    &            & 98393.eth &            & 0908888888.bitcoin & 181818181818.blockchain \\ \cline{2-6} 
    &            & 98441.eth &            & 0908888888.dao     & 1818181818.nft          \\ \cline{2-6} 
    &            & 98094.eth &            & 0998888888.dao     & 181818181818.nft        \\ \hline
  \textbf{\begin{tabular}[c]{@{}c@{}}Total\\ Domains\\ Owned\end{tabular}} &
    3 &
    179 &
    2 &
    13 &
    14 \\ \hline
  \end{tabular}}
  \caption{Examples of targets, typosquatting names, and typosquatting addresses of \textbf{Type (iii)} typosquatting attacks (i.e. where attackers register $>$1 domains targetting similar legitimate domains)}~\label{tab:camp3}
\end{table*}

\begin{figure}[h]
  \centering
      \includegraphics[scale=0.3,frame]{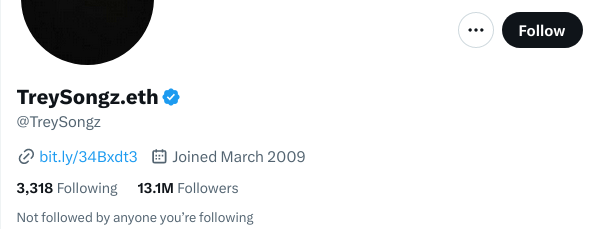} 
      \caption{Profile of the most popular Twitter/X user advertising his ENS domain name.}
      \label{fig:treysongz}
\end{figure}

\begin{figure}[h]
  \centering
  \includegraphics[scale=0.55]{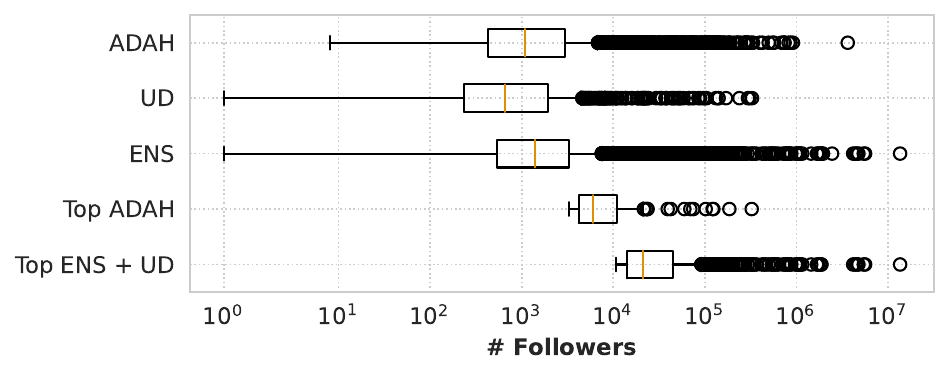}
  \caption{Distribution of the number of followers of all the collected \bns{} names popularized on Twitter/X}\label{twtfollowers}
\end{figure}

\begin{figure}[h]
\centering
\includegraphics[scale=0.55]{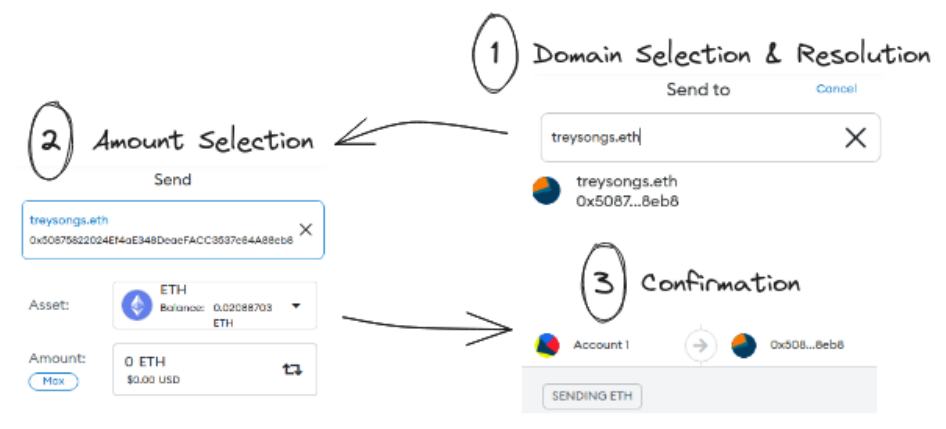}
\caption{Example of the ``Cold Typos'' experiment, showing the stages to sending assets to \texttt{treysongs.eth} (targeting \texttt{treysongz.eth}) through Metamask without warnings.}\label{croppedssmetamask}
\end{figure}

\end{appendices}
\end{document}